\documentclass{article}
\usepackage{spconf,graphicx}
\usepackage{amsmath,amssymb,amsfonts}
\usepackage{graphicx} 
\usepackage{subfigure}
\usepackage{bm}
\usepackage{cite}
\usepackage{flushend}
\usepackage{amsfonts}
\usepackage{stfloats} 

\usepackage{algorithm}
\usepackage{algpseudocode}

\usepackage{booktabs}
\usepackage{tabularx}
\usepackage{amssymb}
\usepackage{graphicx}
\usepackage{threeparttable}
\usepackage{makecell}
\usepackage{extarrows}
\usepackage{epstopdf}
\usepackage{bbding}
 
\usepackage{tabularx}
\usepackage{cuted}
\usepackage{extarrows}
\usepackage{color}

\title{ \vspace{-1.5em} Sensing With Random Signals}

\name{Shihang Lu$^{\dagger}$, Fan Liu$^{\dagger}$, Fuwang Dong$^{\dagger}$, Yifeng Xiong$^{\ddagger}$, Jie Xu$^{\sharp}$, and Ya-Feng Liu$^{\natural }$} 

\address{\textit{(Invited Paper)} \\
$^{\dagger}$ DEEE, Southern University of Science and Technology, China \\
$^{\ddagger}$ SICE, Beijing University of Posts and Telecommunications, China \\
$^{\sharp}$  SSE and FNii, The Chinese University of Hong Kong (Shenzhen), China\\
$^{\natural}$  ICMSEC, AMSS, Chinese Academy of Sciences, China
\thanks{\textit{Corresponding author}: Fan Liu (liuf6@sustech.edu.cn).}}

\usepackage[font=small,skip=0pt]{caption}
\begin{document}
\maketitle

\begin{abstract}
Radar systems typically employ well-designed {\it deterministic} signals for target sensing. In contrast to that, integrated sensing and communications (ISAC) systems have to use {\it random} signals to convey useful information, potentially causing sensing performance degradation. In this paper, we define a new sensing performance metric, namely, ergodic linear minimum mean square error (ELMMSE), accounting for the randomness of ISAC signals. Then, we investigate a data-dependent precoding scheme to minimize the ELMMSE, which attains the optimized sensing performance at the price of high computational complexity. To reduce the complexity, we present an alternative data-independent precoding scheme and propose a stochastic gradient projection (SGP) algorithm for ELMMSE minimization, which can be trained offline by locally generated signal samples. Finally, we demonstrate the superiority of the proposed methods by simulations.
\end{abstract}

\begin{keywords}
Integrated sensing and communications (ISAC), deterministic-random tradeoff,  random signals. 
\end{keywords}
\vspace{-0.5em}
\section{Introduction}
\vspace{-0.5em}
Recently, the International Telecommunication Union (ITU) has recognized integrated sensing and communications (ISAC) as one of the six key  {usage scenarios} for 6G, marking the continuous progression of ISAC from concept to practice \cite{ITU2023}. By reusing wireless resources, ISAC may deploy ubiquitous sensing ability over existing networking infrastructures in a low-cost manner \cite{cui2021integrating,liu_an2022survey,lu2023integrated}. Among various approaches to implement ISAC, utilizing communication signals for sensing holds the most promising potential \cite{sturm2011waveform,lixinyang2023optimal,zhangyumeng2023input}. To convey communication information, ISAC signals should be {\it random}, e.g., capacity-achieving Gaussian signals \cite{cover1999elements,xielei2023sensing}. Conversely, radar systems prefer {\it deterministic} signals, e.g., constant-modulus signals, which may offer improved sensing performance. This contradictory conflict  {leads to the so-called} {\it deterministic-random tradeoff (DRT)} in ISAC systems \cite{xiong2023fundamental}. 

Despite the random nature of ISAC signals, most of previous studies aiming for ISAC signal design treat the sample covariance matrix of transmit signals as if they were deterministic \cite{liu2021cramer,liu_Yafeng2022joint,lu2022performance}, under the assumption that the data frame is sufficiently long. Those approaches overlooked the DRT, making them akin to infinite-frame-length approximations. However, such approximations may not be reliable in practical scenarios. For instance, in massive multiple-input multiple-output (MIMO) systems, the base station may be equipped with massive antennas, and the signal sample covariance matrix converges to its statistical counterpart only when the frame length is much larger than the antenna number. To alleviate the resultant huge computational and signal processing overheads, one has to perform sensing by leveraging short frames, where the law of large numbers no longer holds and the data randomness cannot be neglected . This motivates our investigation in this paper.

Specifically, we consider {\it sensing with random signals} in MIMO ISAC systems, where the transmitter aims to estimate the target impulse response (TIR) matrix by using random information-carrying communication signals. To characterize the sensing performance  {in this practical scenario}, we define a new performance metric, namely, ergodic linear minimum mean-square-error (ELMMSE), which is defined as the average estimation error over  {random} signal realizations. Then, we reveal the performance disparity between deterministic and random signals for sensing, drawing on Jensen's inequality. Subsequently, we present a data-dependent precoding strategy and a stochastic gradient projection (SGP) algorithm to minimize the ELMMSE. Finally, our simulations underscore the performance degradation linked to random signals in sensing, and offer guidance for practical design in scenarios with limited coherent frame lengths.
\vspace{-1.0em}
\section{System Modeling} \label{sec2}
\vspace{-0.5em}

We consider a MIMO ISAC system with the base station (BS) deployed with $N_t$ transmit antennas and $N_r$ receive antennas, where the target sensing is implemented over a  coherent processing interval consisting of $L$ snapshots in a monostatic setup. The echo signals at the ISAC receiver are denoted by 
\vspace{-0.5em}
\small\begin{align}\label{sensing_signals}
    \bm{Y}_{s}& = \bm{H}_s \bm{X}  + \bm{Z}_s,
\end{align}\normalsize
where $\bm{X} \in \mathbb{C}^{N_t \times L }$ is the sensing signal matrix, $\bm{H}_s \in \mathbb{C}^{N_r \times N_t}$ is the TIR matrix to be estimated, and $\bm{Z}_s \in \mathbb{C}^{N_r \times L}$ denotes an additive noise matrix with each entry following $\mathcal{CN}(0,\sigma_s^2)$. In what follows, we commence by sensing with deterministic signals.

\vspace{-0.5em}
\subsection{Sensing With Deterministic Training Signals}
Let $\bm{S}_D \in \mathbb{C}^{N_t \times L}$ denote a {\it deterministic} training signal satisfying $(1/L)\bm{S}_D\bm{S}_D^H = \bm{I}_{N_t} $. The sensing signal matrix is $\bm{X} = \bm{W} \bm{S}_D$, where $\bm{W}\in \mathbb{C}^{N_t \times N_t}$ represents the precoding matrix to be optimized. Then, the celebrated linear minimum mean-square-error (LMMSE) estimation of $\bm{H}_s$ is
 \small\begin{align}\label{Estimator_LMMSE}
\hat{\bm{H}}_{\mathsf{LMMSE}}=\bm{Y}_s\left(\bm{X}^H \bm{R}_H\bm{X}+\sigma_s^2 N_r \bm{I}_{L}\right)^{-1} \bm{X}^H \bm{R}_H,
\end{align}\normalsize 
where $\bm{R}_H = \mathbb{E}[\bm{H}_s^H \bm{H}_s]$ represents the statistical correlation matrix of the channel  { with $\mathbb{E}[\cdot]$ denoting the statistic expectation}. The resulting estimation error is \cite{biguesh2006training}
 \small\begin{align}\label{Cond_MMSE}
{J}_{\mathsf{LMMSE}} 
= \mathrm{tr}\Big[\Big(\bm{R}_H^{-1} + \frac{L}{\sigma_s^2 N_r} \bm{W}\bm{W}^H\Big)^{-1}\Big].
\end{align}\normalsize  
Accordingly, the LMMSE-oriented design is to solve the 
{\it deterministic} optimization problem:
 \small\begin{align}\label{LMMSE_Determinstic}
    \min_{\bm{W}} ~ J_{\mathsf{LMMSE}} ~~
		\mathrm{s.t.} ~ \bm{W} \in \mathcal{A},
\end{align}\normalsize 
where $\mathcal{A} = \{\bm{W} \in \mathbb{C}^{N_t \times N_t}  |~ \| \bm{W} \|_F^2 \le P \}$ denotes the feasible region and $P$ is the power budget. Let $\bm{R}_H = \bm{Q}\bm{\varLambda}\bm{Q}^H$ denote the eigenvalue decomposition of $\bm{R}_H$. The optimal precoding matrix is known to be the following {\it water-filling} solution \cite{biguesh2006training}:
 \small\begin{align}\label{LMMSE_Determinstic_Opt_W}
    \bm{W}_{\mathsf{LMMSE}}^{\mathsf{opt}} = \sqrt{\frac{\sigma_s^2{N_r}}{L}} \bm{Q}\Big[ \Big(\mu_0 \bm{I}_{N_t} - \bm{\varLambda}^{-1}\Big)^{+} \Big]^{\frac{1}{2}},
\end{align}\normalsize 
where $(x)^{+} = \max (x,0)$ and $\mu_0$ is a constant that is determined based on $\|\bm{W}_{\mathsf{LMMSE}}^{\mathsf{opt}}\|_F^2 = P$.  
\vspace{-0.5em}

\subsection{Sensing With Random Communication Signals}
In contrast to conventional radar systems, ISAC systems have to employ random communication signals for sensing. In this paper, we consider random signaling for ISAC systems. Let $\bm{S} = [\bm{s}(1), \dots, \bm{s}(L)] \in \mathbb{C} ^{N_t \times L}$ denote the transmitted ISAC signal, where each column is independent and identically distributed (i.i.d.) with zero mean and covariance $\bm{I}_{N_t}$. It is worth noting that the following assumption is commonly seen in the existing works \cite{liu2021cramer,liu_Yafeng2022joint,lu2022performance},
 \small\begin{align}\label{SampleMatrix_Appro}
    \frac{1}{L} \bm{S}\bm{S}^H \approx \mathbb{E}[\bm{s}(l)\bm{s}^H(l)] = \bm{I}_{N_t}, l = 1,\dots,L, 
\end{align}\normalsize 
where $L \gg N_t$  {is assumed} such that the approximation holds. However, this assumption cannot always be satisfied, especially when the BS does not have the ability to accumulate sufficiently long frames. To this end, one needs to take the randomness of the ISAC signals into account when designing the precoding matrix. More severely, the sensing performance may be randomly varying due to the random signaling. As a consequence, it may not be appropriate to use the conventional estimation metric relying on the instantaneous realization of signals. Therefore, we define a new {\it ELMMSE} to measure the average sensing performance over random signal realizations, which is expressed as 
 \small\begin{align}
    {J}_{\mathsf{ELMMSE}} = \mathbb{E}_{\bm{S}} \left\{\mathrm{tr}\Big[(\bm{R}_H^{-1} +\frac{1}{\sigma_s^2 N_r} \bm{W}\bm{S}\bm{S}^H\bm{W}^H)^{-1}\Big] \right\}.
\end{align}\normalsize 
In this paper, we aim to minimize the average sensing performance under random signaling, which is modeled by \footnote{In general, $\bm{W}$ can be designed depending on random signal realization $\bm{S}$. This thus leads to two different designs of $\bm{W}$, which remains unchanged and changes adaptively based on the instantaneous value of $\bm{S}$, referred to as data-independent and data-dependent designs in our paper, respectively.}
 \small\begin{align}\label{LMMSE_op}
		\min_{\bm{W}} ~~ f(\bm{W}) \triangleq J_{\mathsf{ELMMSE}} 
		~~ \mathrm{s.t.} ~ \bm{W} \in \mathcal{A}.
\end{align}\normalsize 

{\it Proposition 1:} Random ISAC signals lead to ELMMSE that is no lower than the LMMSE based on the deterministic signals, i.e., LMMSE serves as a lower bound of ELMMSE.

{\it Proof:}
    By applying the Jensen's inequality to ELMMSE, we have the following relationship
     \small\begin{align}\label{Jensen}
    {J}_{\mathsf{ELMMSE}} & \overset{(a)}{\ge} \mathrm{tr}\Big[\big(\bm{R}_H^{-1} + \frac{1}{\sigma_s^2 N_r} \mathbb{E}_{\bm{S}}\big\{\bm{WS}\bm{S}^H\bm{W}^H\big\}\big)^{-1}\Big] \nonumber\\
    &\overset{(b)}{=} \mathrm{tr}\Big[\big(\bm{R}_H^{-1} + \frac{1}{\sigma_s^2 N_r}\bm{X}\bm{X}^H\big)^{-1}\Big]{=} {J}_{\mathsf{LMMSE}},
    \end{align}\normalsize 
    where $(a)$ holds due to the convexity of $ {J}_{\mathsf{ELMMSE}}$ with respect to $\bm{WS}\bm{S}^H\bm{W}^H$, and $(b)$ holds when $\bm{S}$ or $\bm{WS}$ is deterministic. This completes the proof. \hfill $\square$

\begin{figure}[htbp]
	\centering
	\includegraphics[scale=0.32]{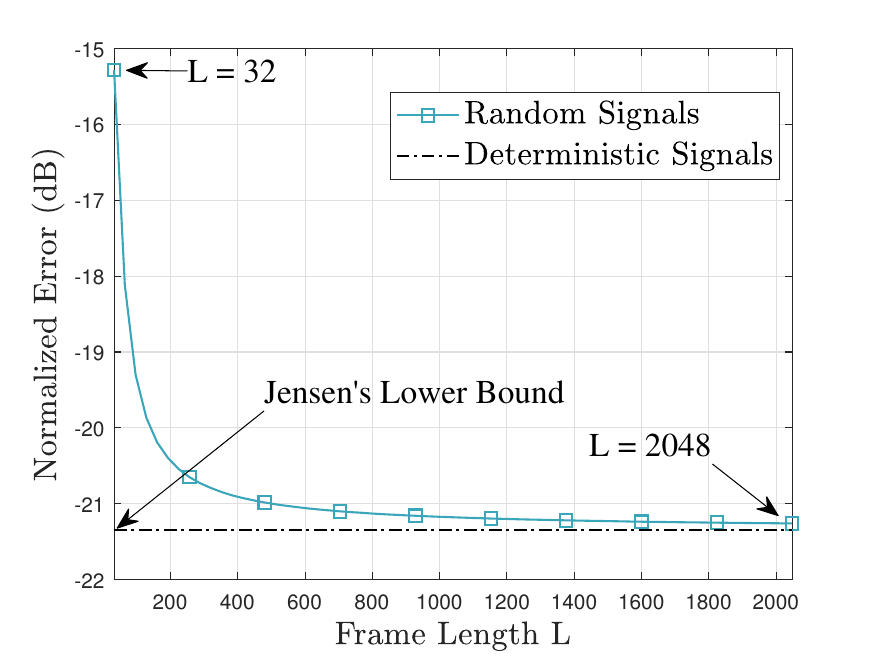}\\
	\caption{The asymptotic performance of random signals.}
        \label{IncreasingL}
\end{figure}

{\it Remark 1:} We notice that the Jensen's inequality in \eqref{Jensen} tends to be tight with $L$ going to infinity. In light of Proposition 1, we show the asymptotic performance of random signals in Fig. \ref{IncreasingL}, precoded by the water-filling precoder given in \eqref{LMMSE_Determinstic_Opt_W}, indicating that the assumption in \eqref{SampleMatrix_Appro} is not always reliable. It also justifies the use of ELMMSE instead of LMMSE as  {the performance metric or design objective} under random signaling, since minimizing the LMMSE  {only minimizes} the lower bound of the ELMMSE  {instead of the exact value}. This observation motivates us to present efficient ELMMSE-oriented precoding solutions for solving problem \eqref{LMMSE_op} in the next section.

\vspace{-1.2em}
\vspace{-2.5em}

\section{ELMMSE-Oriented precoding}\label{Sec3}
\vspace{-0.5em}
\subsection{Data-Dependent Precoding}
\vspace{-0.5em}

We first take the stochastic approximation into account by $N$ samples of random ISAC signals, written as 
 \small\begin{align}\label{Sample_Appro_LMMSE}
    {f}(\bm{W}) \approx \frac{1}{N} \sum_{n=0}^{N-1} \hat{f}(\bm{W};\bm{S}_n), 
\end{align}\normalsize 
where $\hat{f}(\bm{W};\bm{S}_n) = \mathrm{tr}[(\bm{R}_H^{-1} + \frac{1}{\sigma_s^2 N_r} \bm{W}\bm{S}_n\bm{S}_n^H\bm{W}^H)^{-1}]$ and $\bm{S}_n$ denotes the $n$-th transmitted data realization. Observe that each $\bm{S}_n$ is known to both the transmitter and sensing receiver due to the monostatic sensing setup. Then the precoding matrix $\bm{W}$ can be seen as a function of $\bm{S}$ over different data realization indices $n$, written as $\bm{W}(\bm{S}_n)$. Therefore, based on \eqref{Sample_Appro_LMMSE}, problem \eqref{LMMSE_op} is approximated as
\vspace{-0.5em}
 \small\begin{align}\label{LMMSE_Sam}
		\min_{\{ \bm{W}(\bm{S}_n)\}_{n=1}^{N}} ~ \frac{1}{N} \sum_{n=0}^{N-1} \hat{f}(\bm{W}(\bm{S}_n); \bm{S}_n) 
		 ~~\mathrm{s.t.} ~\bm{W}(\bm{S}_n ) \in \mathcal{A}.
\end{align}\normalsize
Consequently, the precoding matrix can be optimized successively with given $\bm{S}_n$ over different data realization $n$, which is called data-dependent precoding. Then this problem is decomposed following $N$ parallel deterministic sub-problems:
\vspace{-0.5em}
 \small\begin{align}\label{LMMSE_SubProblem}
		\min_{\bm{W}_n} ~ \hat{f}(\bm{W}_n; \bm{S}_n) 
		 ~~\mathrm{s.t.} ~\bm{W}_n \in \mathcal{A}, 
\end{align}\normalsize 
where we define $\bm{W}_n \triangleq \bm{W}(\bm{S}_n)$ for notational convenience. Although problem \eqref{LMMSE_SubProblem} is non-convex, we propose to solve it using the successive convex approximation (SCA) algorithm \footnote{We have derived the optimal solution with closed form in \cite{lu2023random} and readers are referred to \cite[Sec. III-A]{lu2023random} for details.}. Let us first approximate the objective function by its first-order Taylor expansion as
\vspace{-0.5em}
 \small\begin{align}
    \hat{f}(\bm{W}_n; \bm{S}_n) \approx \hat{f}( &\bm{W}_n^{(t)}; \bm{S}_n) \nonumber \\
    &+  \langle\nabla {\hat{f}} (\bm{W}_n^{(t)}; \bm{S}_n) ,(\bm{W}_n -\bm{W}_n^{(t)}) \rangle,
\end{align}\normalsize 
where $\langle\bm{A},\bm{B} \rangle$ represents the inner product of $\bm{A}$ and $\bm{B}$, $\bm{W}_n^{(t)} \in \mathcal{A}$ denotes a given feasible point, and $\nabla {\hat{f}}(\cdot)$ represents the gradient of $\hat{f}(\cdot)$, calculated by

 \small\begin{align}
\nabla\hat{f}(\bm{W}_n^{(t)};\bm{S}_{n}) =- \frac{1}{\sigma_s^2 N_r} \bm{A}_n^{-2}\bm{W}_n^{(t)} \bm{S}_n\bm{S}_n^H, 
\end{align}\normalsize 
where $\bm{A}_n = \bm{R}_H^{-1} + \frac{1}{\sigma_s^2 N_r} \bm{W}_n^{(t)}\bm{S}_n\bm{S}_n^H(\bm{W}_n^{(t)})^H$.

At the $t$-th iteration of the SCA algorithm, we solve the following optimization problem:

\begin{align}\label{LMMSE_SCA}
		\min_{\bm{W}_n} & ~~ {g}(\bm{W}_n) \triangleq  \langle\nabla {\hat{f}} (\bm{W}_n^{(t)}; \bm{S}_n) ,(\bm{W}_n -\bm{W}_n^{(t)}) \rangle \nonumber \\ 
		\mathrm{s.t.}& ~~ \bm{W}_n \in \mathcal{A}.
\end{align}\normalsize 
We observe that problem \eqref{LMMSE_SCA} is convex and can be solved by off-the-shelf numerical tools  {like CVX} \cite{boyd2004convex}. We denote the solution to problem \eqref{LMMSE_SCA} as $\bm{W}_n^{\prime} \in \mathcal{A}$. We observe that ${g}(\bm{W}_n^{\prime}) \le 0$, since ${g}(\bm{W}_n^{(t)})  = 0$, which indicates that $(\bm{W}_n^{\prime} - \bm{W}_n^{(t)})$ represents a descent direction for the $(t+1)$-th iteration. Therefore, we can update the precoding matrix by moving along this direction with a certain step size $\delta^{(t)}$, i.e.,
 \small\begin{align}
    \bm{W}_n^{(t+1)} = \bm{W}_n^{(t)} + \delta^{(t)}(\bm{W}_n^{\prime}-\bm{W}_n^{(t)}), ~\delta^{(t)} \in [0,1]. 
\end{align}\normalsize 

The SCA algorithm for solving \eqref{LMMSE_SubProblem} is summarized in \textbf{Algorithm} \ref{SCA_Alg}. By solving a series of parallel sub-problems \eqref{LMMSE_SubProblem}, we obtain the set of data-dependent precoding matrix tailored for each $\bm{S}_n$, denoted by $\mathcal{W} = \{ \bm{W}_1^{\star}, \dots,\bm{W}_N^{\star}\}.$ Accordingly, the ELMMSE may be calculated as
\vspace{-0.5em}
\small\begin{align}
{J}_{\mathsf{ELMMSE}}  = \frac{1}{N} \sum_{n=0}^{N-1} \hat{f}(\bm{W}^{\star}_n,\bm{S}_n).
\end{align}\normalsize

{\it Remark 2:} The data-dependent precoding represents the performance lower bound of \eqref{LMMSE_Sam}. Notice that $\mathcal{W}$ needs to be designed sequentially for different realizations of transmitted random signals. This, however, results in high computational complexity in practical scenarios. This motivates us to harness the popular SGP algorithm with lower complexity, where a single data-independent precoding matrix is employed for all data realizations, as detailed below. 
\vspace{-0.5em}
\begin{algorithm}[h]
    \caption{SCA Algorithm for Solving     \eqref{LMMSE_SubProblem}.}
    \label{SCA_Alg}
    \begin{algorithmic}[1]
    \Require
    $\bm{S}_n, \bm{R}_H, P, N, \sigma^{2}_{s} , N_t,N_r, L, t_{\mathsf{max}}, \xi < 0 $.
    \Ensure
    $\bm{W}^{\star}_n = \bm{W}_n^{(t)}$.
    \State Initialize the precoding matrix $\bm{W}_n^{(t)}, r = t$.
    \State Initialize the random ISAC signal set $\mathcal{S} = \{\bm{S}_1, \ldots\, \bm{S}_N\}$.
    \Repeat
    \State Solve the convex Problem \eqref{LMMSE_SCA} to obatin $\bm{W}_n^{\prime} \in \mathcal{A}$ .
    \State Update $\bm{W}_n^{(t+1)} = \bm{W}^{(t)} + \delta^{(t)}(\bm{W}_n^{\prime}-\bm{W}^{(t)})$, where the step $\delta^t$ can be set by applying the exact line search.
    \State Update $t=t+1$.
    \Until ${g}(\bm{W}_n^{(t)}) \le \xi $ or $t=t_{\mathsf{max}}$.
    \end{algorithmic}
\end{algorithm}
\vspace{-0.5em}

\vspace{-0.5em}
\begin{algorithm}[htbp]
    \caption{SGP Algorithm for Solving \eqref{LMMSE_op}.}
    \label{PSGD_Alg}
    \begin{algorithmic}[1]
    \Require
    $\bm{R}_H, P, N, \sigma^{2}_{s} , N_t,N_r,r_{\mathrm{max}},\epsilon$.
    \Ensure
    $\bm{W}$.
    \State Initialize the precoding matrix $\bm{W}^{(r)}, r = 1$.
    \State Initialize the random signal set $\mathcal{S} = \{\bm{S}_1, \ldots\, \bm{S}_N\}$.
    \Repeat
    \State Generate $\mathcal{D}^{(r)}$ and calculate $\hat{\nabla}{f}(\bm{W}^{(r)},\bm{S}^{(r)})$.
    \State Update $\bm{W}^{(r+1)} = \mathsf{Proj}_{\mathcal{A}}\Big(\bm{W}^{(r)} - \eta^{(r)} \hat{\nabla}{f}(\bm{W}^{r})\Big)$.
    \State Update $r=r+1$.
    \Until The increase of the objective value is below $\epsilon$ or $r=r_{\mathrm{max}}$.
    \end{algorithmic}
\end{algorithm}
\vspace {-2.0em}

\subsection{SGP-Based Data-Independent Precoding}
To seek for a data-independent precoder, we resort to the stochastic gradient descent (SGD) algorithm, which yields a solution with only one or mini-batch samples in each iteration, reducing computational complexity. At the $r$-th iteration, the precoding matrix $\bm{W}$ is updated as
 \small\begin{align}\label{BF_SGDupdate}
 \bm{W}^{(r+1)} \leftarrow \bm{W}^{(r)} - \eta^{(r)} \hat{\nabla} {f}(\bm{W}^{(r)}),
\end{align}\normalsize 
where $\bm{W}^{(r)}$ denotes the precoding matrix in iteration $r$, $\eta^{(r)}$ represents the step size, and $\hat{\nabla} {f}(\bm{W}^{(r)})$ is the gradient obtained from a local randomly generated mini-batch sample set $\mathcal{D}^{(r)}$, calculated by 
\vspace{-0.5em}
\small\begin{align}
\hat{\nabla} {f}(\bm{W}^{(r)}) = \frac{1}{|\mathcal{D}^{(r)}|} \sum_{\bm{S} \in \mathcal{D}^{(r)}} {\nabla} \hat{f}(\bm{W}^{(r)}; \bm{S}).
\end{align}\normalsize 

In order to adhere to the transmit power budget, we propose to utilize the SGP algorithm, where we project $\bm{W}^{(r+1)}$ onto the convex feasible region $\mathcal{A}$, i.e.,
\vspace{-0.5em}
\small\begin{align}\label{Proj}
\mathsf{Proj}_{\mathcal{A}}( \bm{W})=\left\{\begin{array}{l}
            \bm{W},~ \mathrm{if}~\bm{W} \in \mathcal{A}, \\
            \bm{W}\sqrt{\frac{P}{\|\bm{W}\|_F^2}},~ \mathrm{Otherwise}. 
        \end{array}\right.
\end{align}\normalsize 
Following the iteration format in \eqref{BF_SGDupdate} and the projection step in \eqref{Proj}, we are now ready to introduce the proposed SGP algorithm to solve \eqref{LMMSE_op}, as detailed in \textbf{Algorithm} \ref{PSGD_Alg}.
 
\begin{table}[htbp]
	\centering
	\fontsize{9.0}{8.0}\selectfont
	\caption{Parameters in Simulations.}\label{tab1}
		{\begin{tabular}{p{1.0cm} p{1.0cm} p{1.0cm} p{1.0cm}l p{1.0cm} p{1.0cm}}
				\hline 
				\hline 
				Parameter & Value & Parameter & Value & Parameter & Value \\
				\hline
				$N_t$ & $64$ &	$N_r$ & $32$ & $N$ & $100$ \\				
				$\sigma_s^2$	& $0~\mathrm{dBm}$ & $P$	& $30~\mathrm{dBm}$  &  $\xi$	& $-0.1$ \\
				$t_{\mathsf{max}}$ & $30$ &	$r_{\mathsf{max}}$ & $2000$	& $\epsilon$  & $10^{-5}$  \\			
				\hline 
		\end{tabular}}
\end{table}
\vspace{-1.5em}

\section{Simulation Results}
\vspace{-0.5em}
In this section, we demonstrate the numerical results of the proposed methods. If not otherwise specified, the transmit SNR is defined as $LP/ \sigma_s^2$. The simulation parameters are listed in Table \ref{tab1}. For the SGP method, we set the number of mini-batch samples as $|\mathcal{D}^{(r)}| = 10$. The eigenvalues of $\bm{R}_H$ represent the spatial channel correlation, which follow a uniform distribution on the interval $[1,10]$ in the simulation.
\vspace{-1em}
\begin{figure}[htbp]
\centering
\subfigure[SCA Algorithm.]{
\includegraphics[scale=0.275]{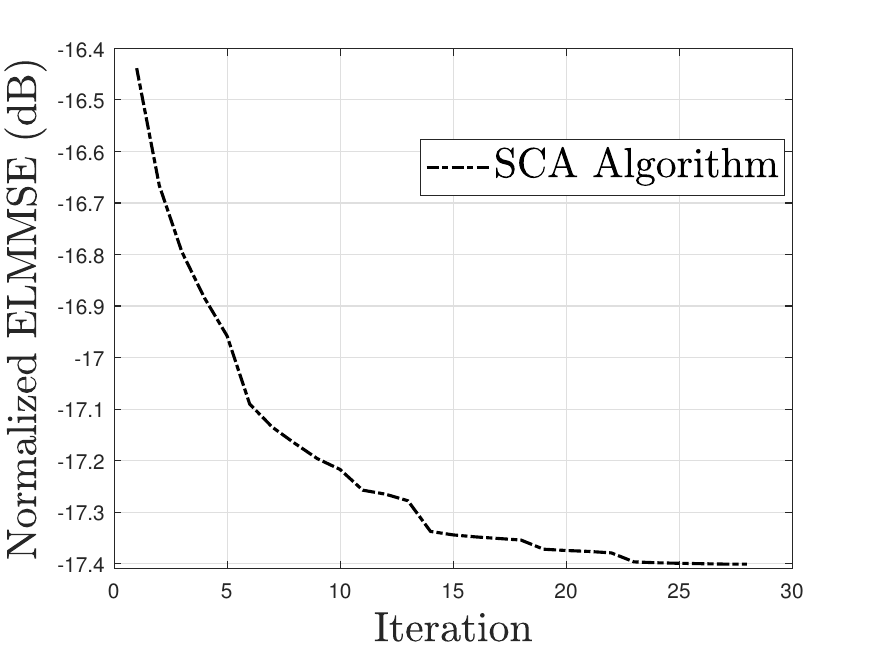} \label{Convergence_SCA} 
}
\subfigure[SGP Algorithm.]{
\includegraphics[scale=0.275]{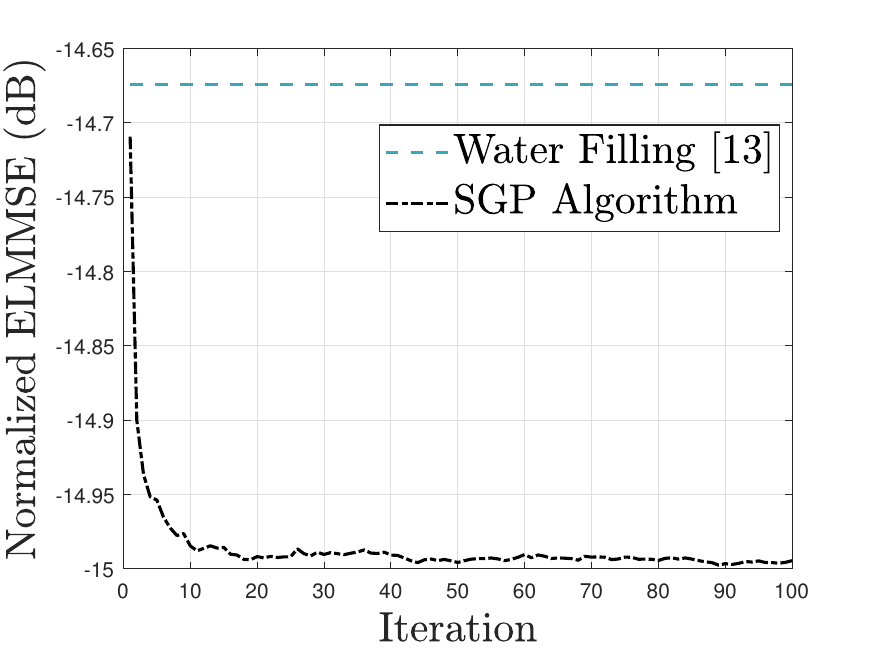} \label{Convergence_SGP}
}
\caption{Convergence examples of proposed algorithms.}
\label{Convergence}
\end{figure}
\vspace{-2em}

In Fig. \ref{Convergence}, we set the frame length $L = 32$ and transmit SNR $=30~\mathrm{dB}$ to show the convergence of our proposed algorithms in Section \ref{Sec3}. In Fig. \ref{Convergence_SCA}, we illustrate the convergence of our proposed SCA algorithm for one of the parallel sub-problems \eqref{LMMSE_SubProblem} within 30 iterations. In Fig. \ref{Convergence_SGP}, we show the convergence of our proposed SGP algorithm, which can be implemented {\it offline} based on the locally generated random signal samples, and the step size is set as $\eta^{(r)} = 10/(10+r)$ to guarantee the convergence \cite{eon1998online}. It is observed that SGP converges quickly within tens of iterations. 

\begin{figure}[htbp]
\centering
\subfigure[$L = 32, N_t = 64$.]{
\includegraphics[scale=0.275]{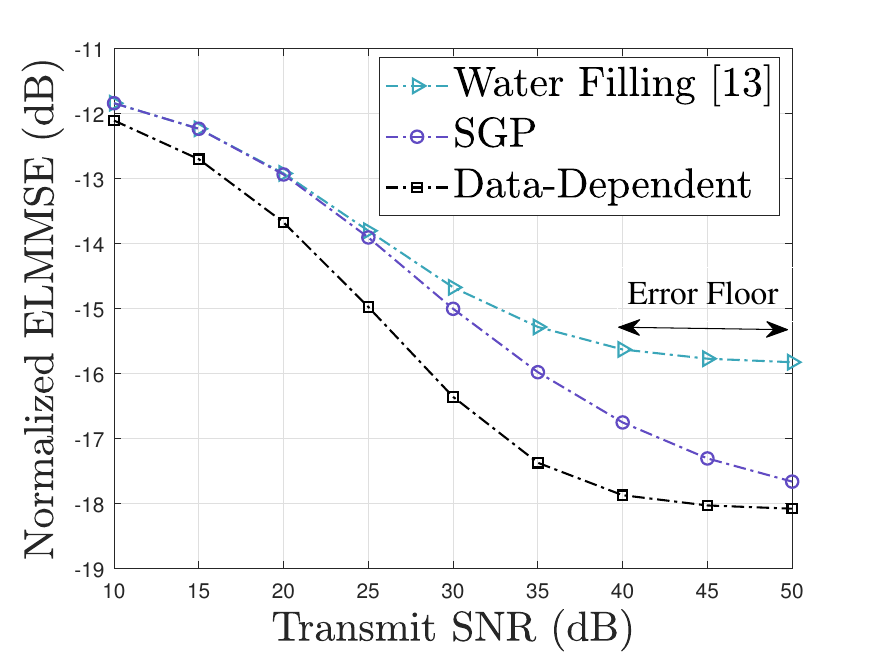} \label{L32Nt64}
}
\subfigure[$L = 64, N_t = 64$.]{
\includegraphics[scale=0.275]{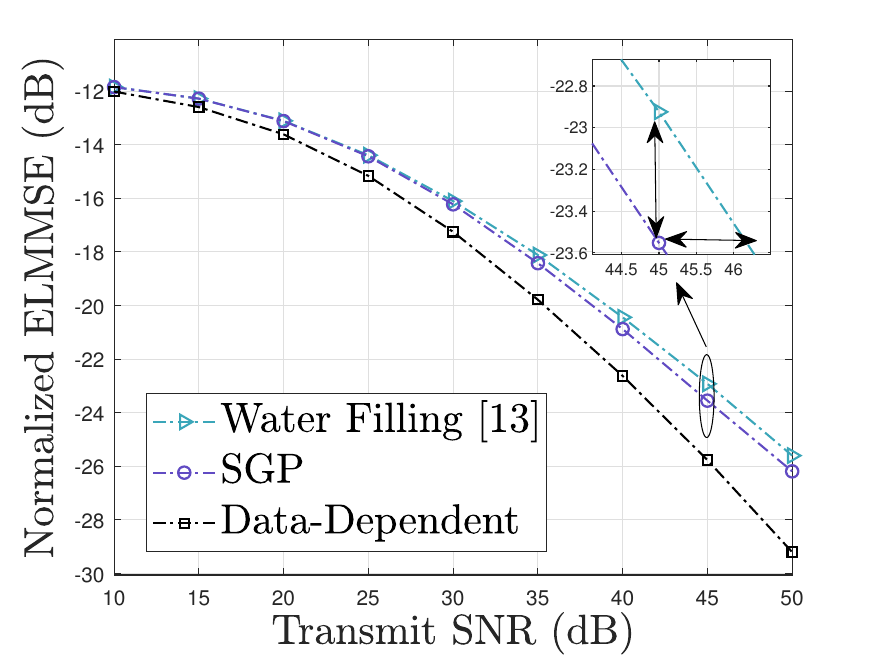} \label{L64Nt64} 
}\caption{Performance comparison among different schemes versus transmit SNR.}
\label{Performance}
\end{figure}

In Fig. \ref{Performance}, we compare three precoding schemes, namely, classical water-filling, SGP, and data-dependent precoding. Notably, the SGP algorithm outperforms the classical water-filling scheme in both cases with $L = 32$ and $L = 64$. This is attributed to two key factors. When $L = 32$, the rank-deficient nature of the signal (see Fig. \ref{L32Nt64}) causes an error floor in the water-filling scheme, rendering it suboptimal even for deterministic signals. Secondly, the inherent randomness of signals in ISAC scenarios is disregarded by the classical water-filling scheme. Moreover, the data-dependent precoding achieves the best performance, leveraging precise knowledge of transmitted data. Nevertheless, adapting data-dependent precoding for varying data realizations requires complex redesigns, incurring significant complexity. Fortunately, a data-independent precoder can be implemented offline through the SGP, with a favorable performance-complexity tradeoff.

\vspace{-1em}
\vspace{-2em}
\vspace{-1em}
\begin{figure}[htbp]
	\centering
	\includegraphics[scale=0.32]{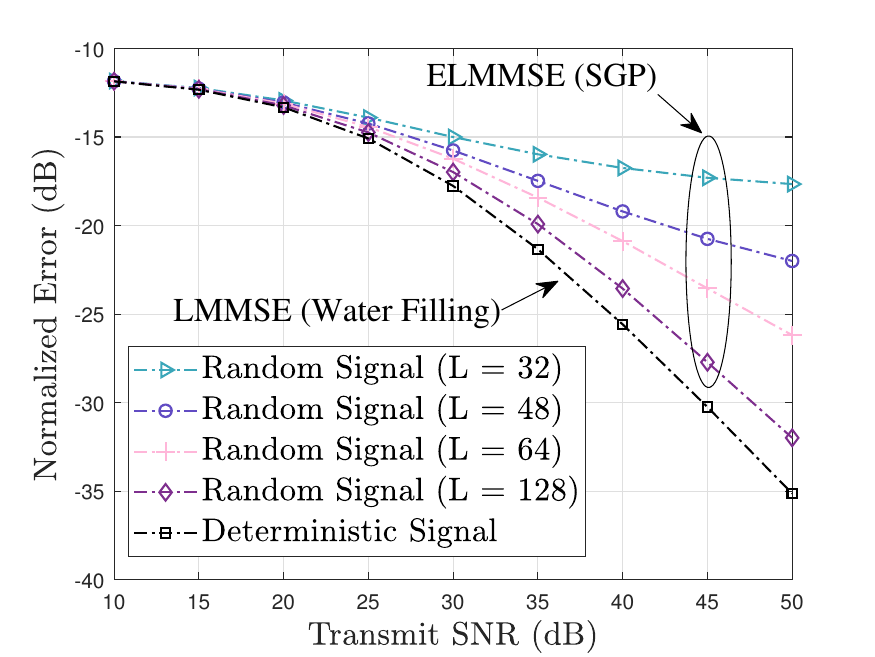}\\
	\caption{Performance comparison between deterministic and random signals versus the transmit SNR.}
        \label{ELMMSE_vs_Jensen}
\end{figure}
\vspace{-1em}

In Fig. \ref{ELMMSE_vs_Jensen}, we show the resultant estimation error by transmitting random (communication) and deterministic (training) signals, with an increasing SNR. All the curves of random signals are obtained by SGP. It is observed that with the increasing frame length, the average sensing performance is asymptotically close to the LMMSE using deterministic signals since the randomness of the signals is decreased. However, there is still a performance gap between using deterministic and random signals, already represented by Jensen's inequality.
\vspace{-1em}

\section{Conclusion}

In this paper, we introduced a new performance metric, namely, ELMMSE, to assess the performance of employing random signals for target sensing. We first revealed that there is a certain performance gap between deterministic and random signals used in sensing, a revelation enabled by Jensen's inequality. Then, we shed light on the precoding matrix design while taking the randomness of the transmit signals into account. To minimize the predefined ELMMSE, we proposed a data-dependent precoding approach and a data-independent SGP scheme, respectively. Finally, simulation results indicated the superiority of our proposed design schemes. In ISAC systems, we ascertain the necessity of accounting for randomness when utilizing random ISAC signals for sensing, particularly for scenarios with short frame lengths.

\bibliographystyle{IEEEtran}
\bibliography{refs}

\end{document}